\documentclass[aps,jap,reprint,showpacs,twocolumn,superscriptaddress]{revtex4-1}

\usepackage[dvips]{graphicx}
\usepackage{dcolumn}
\usepackage{bm}
\usepackage{xspace}
\usepackage{multirow}
\usepackage{natbib}
\usepackage{color}

\renewcommand{\vec}{\mathbf}
\begin{document}

\preprint{}
\title{Absence of a Dirac Gap in Ferromagnetic Cr$_x$(Bi$_{0.1}$Sb$_{0.9}$)$_{2-x}$Te$_3$}
\author{Chung Koo Kim}
\affiliation{Condensed Matter Physics and Materials Science Department, Brookhaven National Lab, Upton, New York 11973, USA\\}
\author{Jonathan D. Denlinger}
\affiliation{Advanced Light Source, Lawrence Berkeley National Laboratory, Berkeley, 94305, California, USA}
\author{Asish K. Kundu}
\author{Genda Gu}
\author{Tonica Valla}
\email{valla@bnl.gov}
\affiliation{Condensed Matter Physics and Materials Science Department, Brookhaven National Lab, Upton, New York 11973, USA\\}

\date{\today}

\begin{abstract}
Magnetism breaks the time reversal symmetry expected to open a Dirac gap in 3D topological insulators that consequently leads to quantum anomalous Hall effect. The most common approach of inducing ferromagnetic state is by doping magnetic 3$d$ elements into bulk of 3D topological insulators. In Cr$_{0.15}$(Bi$_{0.1}$Sb$_{0.9}$)$_{1.85}$Te$_3$, the material where the quantum anomalous Hall effect was initially discovered at temperatures much lower than the ferromagnetic transition, $T_C$, the scanning tunneling microscopy studies have reported a large Dirac gap $\sim20-100$ meV. The discrepancy between the low temperature of quantum anomalous Hall effect ($\ll T_C$) and large spectroscopic Dirac gaps ($\gg T_C$) found in magnetic topological insulators remains puzzling. Here, we used angle-resolved photoemission spectroscopy to study the surface electronic structure of pristine and potassium doped surface of Cr$_{0.15}$(Bi$_{0.1}$Sb$_{0.9}$)$_{1.85}$Te$_3$. Upon potassium deposition, the $p$-type surface state of pristine sample was turned into an $n$-type, allowing spectroscopic observation of Dirac point. We find a gapless surface state, with no evidence of a large Dirac gap reported in tunneling studies.

\end{abstract}
\vspace{1.0cm}


\maketitle

\section{Introduction}

Some of the most exotic electronic phenomena predicted for the surface states of 3D topological insulators (TIs), involve opening the gaps in their spectrum. In one case, opening of a superconducting gap could lead to zero-energy Majorana modes that could serve as a platform for fault-tolerant quantum computing \cite{Fu2008,Qi2011}. The second case involves magnetism and breaking of time-reversal symmetry that is expected to open a Dirac gap and lead to quantum anomalous Hall effect (QAHE) - a dissipationless quantum Hall states in the absence of external magnetic field \cite{Yu2010,Chang2013,He2013}. The QAHE has been recently detected in several magnetic topological materials, but the observation has been limited to sub-Kelvin temperatures, much lower than the magnetic ordering temperature \cite{Chang2013}. The discrepancies are even larger if one considers the Dirac gaps reported by some spectroscopic probes - these are typically several orders of magnitude larger than the range of QAHE existence. Use of magnetic dopant atoms to generate a ferromagnetic state was the first approach to break the time reversal symmetry in existing TIs. The QAHE was initially discovered in a magnetically doped TI - Cr$_x$(Bi$_{0.1}$Sb$_{0.9}$)$_{2-x}$Te$_3$ \cite{Chang2013}. More recently, several intrinsic magnetic TIs have been discovered, ranging from EuSn$_2$P$_2$ and EuSn$_2$As$_2$ \cite{Gui2019,Li2019} to MnBi$_2$Te$_4$ \cite{Otrokov2019,Zeugner2019,Chen2019b,Li2019,Chen2019a} These are stoichiometric layered materials consisting of magnetically ordered layers sandwiched by layers that ensure the TI character of the material. The magnetic layers are usually ferromagnetically ordered but are antiferromagnetically coupled in the crystal. The long range magnetic order is usually established below 20-30 K. The recent studies on MnBi$_2$Te$_4$ report the QAHE at record high temperatures $\sim5$ K with the longitudinal transport gap of $\sim0.6$ meV \cite{Liu2020,Deng2020}. However, the spectroscopic evidence for the Dirac gap in this material has been quite conflicting. In studies where it was detected, it was found to be of the order of 20-100 meV, much higher than magnetic ordering temperature \cite{Otrokov2019,Zeugner2019,Rienks2019,Shikin2020}. Several other studies have found no Dirac gap and no change in the surface electronic structure between the normal and magnetically ordered states, but with some indications of magnetic order in the spectral features of bulk states \cite{Hao2019,Chen2019b,Li2019,Swatek2020,Nevola2020}.\\
The early spectroscopic indicatives of a Dirac gaps on TI surfaces decorated by magnetic overlayers and bulk-doped by magnetic elements came from angle-resolved photoemission spectroscopy (ARPES) \cite{Wray2010a,Chen2010,Xu2012a}. However, these studies were later questioned and required to be re-interpreted as the gaps were present in both the magnetically ordered state and at temperatures high above the ordering transition, as well as in the samples without long range order \cite{Bianchi2011,Valla2012a,Sanchez-Barriga2016}. The STM studies were also inconclusive and in discrepancy with low QAHE and magnetic ordering temperatures. The STM study on Cr$_x$(Bi$_{0.1}$Sb$_{0.9}$)$_{2-x}$Te$_3$, the same material where the QAHE was initially discovered, shows and inhomogeneous Dirac gap, $2\Delta_D$, in the range 20-100 meV whereas the Curie temperature is $T_C=18$ K \cite{Lee2015c}. That material is intrinsically $p$-type, with the Dirac point being unoccupied, preventing the ARPES to probe the effects of magnetic ordering on the Dirac spectrum. In the present study, we use the surface electron doping by depositing potassium to bring the Dirac point below the Fermi level. Our results show gapless, or nearly gapless Dirac excitations, inconsistent with the large Dirac gap reported by STM \cite{Lee2015c}. 

\begin{figure*}[htpb]
\begin{center}
\includegraphics[width=13cm]{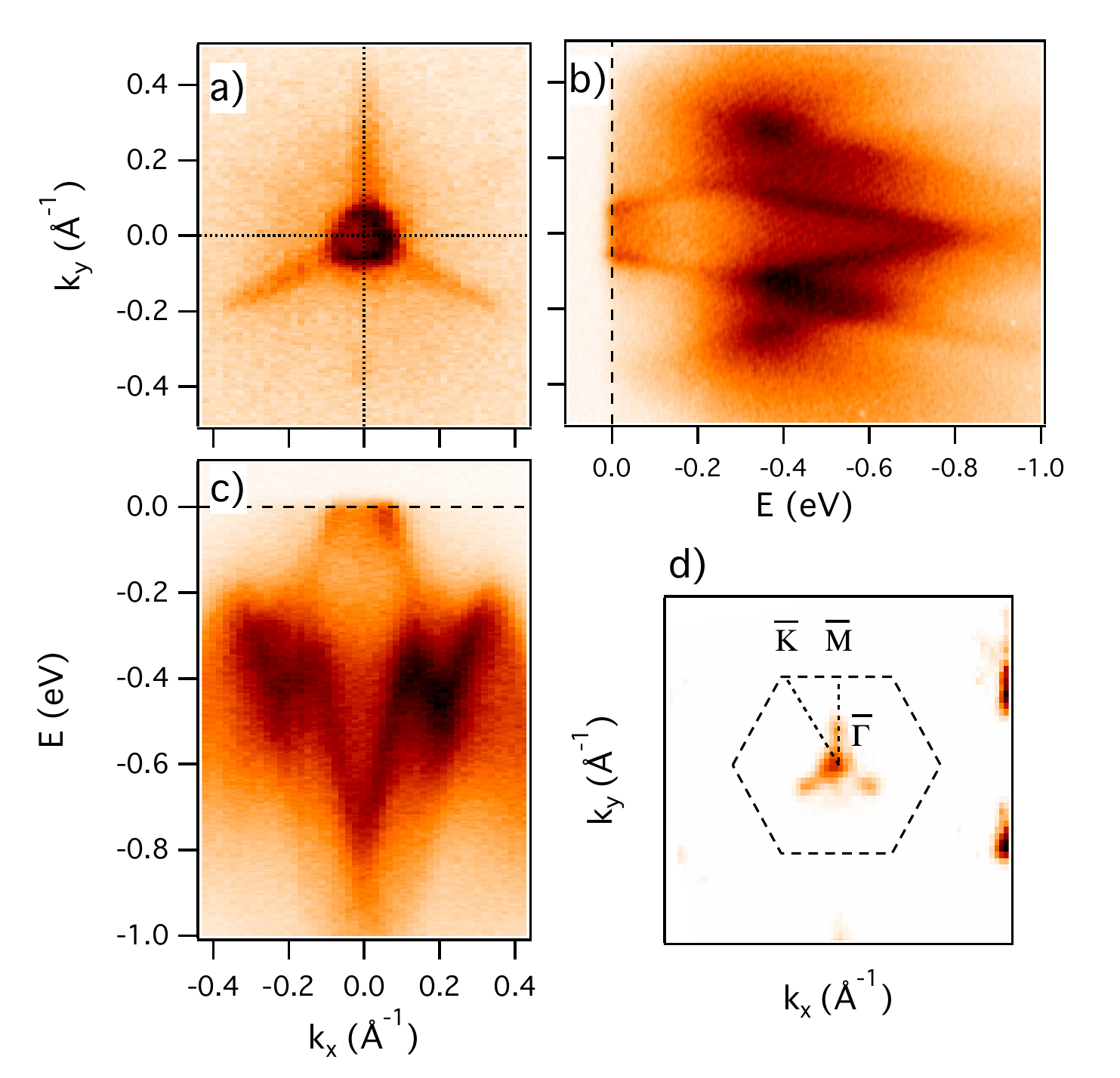}
\caption{Electronic structure of pristine Cr$_{0.15}$(Bi$_{0.1}$Sb$_{0.9}$)$_{1.85}$Te$_3$. (a) Fermi surface. (b) Electronic dispersions along the $\bar{M}-\bar{\Gamma}-\bar{M}$ momentum line. (c) Electronic dispersions along the $\bar{K}-\bar{\Gamma}-\bar{K}$ momentum line. (d) Fermi surface on a large momentum scale. The dotted hexagon represents the first surface Brillouin zone (SBZ). All the spectra are taken at 70 eV photon energy at 12 K. 
}
\label{Fig1}
\end{center}
\end{figure*}
%
\section{Experimental Results}
\subsection{Experimental Methods}
Single crystals with nominal composition Cr$_x$(Bi$_{0.1}$Sb$_{0.9}$)$_{2-x}$Te$_3$ were grown by a modified floating-zone method. The elements of high purity (99.9999\%) Bi, Sb, Cr, and Te were loaded into double-walled quartz ampoules and sealed under vacuum. The materials first were melted at $900^{\circ}$ C in a box furnace and fully rocked to achieve homogeneous mixture. The 12 mm diameter premelt ingot rods in a quartz tube were mounted in a floating-zone furnace. In the floating-zone furnace, the premelt ingot rods were first premelted at a velocity of 200 mm/h and then grown at 1.0 mm/h in 1 bar Ar atmosphere. Because the segregation coefficient of chromium is less than 1, the Cr contained in the feed material would then prefer to remain in the liquid zone. As a result, a homogeneous Cr concentration along the whole grown rod is difficult to achieve. The Cr concentration in the \textit{as-grown} single crystals is thus somewhat less than the nominal Cr-concentration in the feed rod.
Overall magnetic properties of the samples used in this study  were evaluated using SQUID magnetometry. The magnetization exhibit the common ferromagnetic behavior with the bulk Curie temperature $T_C\sim18$ K and coercive field $H_C\sim15$ mT at $T=4.5$ K. 

Photoemission data were collected at Advanced Light Source, at the beamline 4.0.3, employing linearly polarized 70 eV photons and Scienta R8000 analyzer in dithered fixed mode, $\pm15^{\circ}$ angular lens mode, and polar angle scanning in steps of $1/3^{\circ}$ for the perpendicular direction. The sample was attached to the cryostat, cleaved and measured at 12 K. The total instrumental energy resolution was $\sim$ 8 meV. Angular resolution was better than $\sim 0.15^{\circ}$ and $0.4^{\circ}$ along and perpendicular to the slit of the analyzer, respectively. 
\begin{figure*}[htbp]
\begin{center}
\includegraphics[width=12cm]{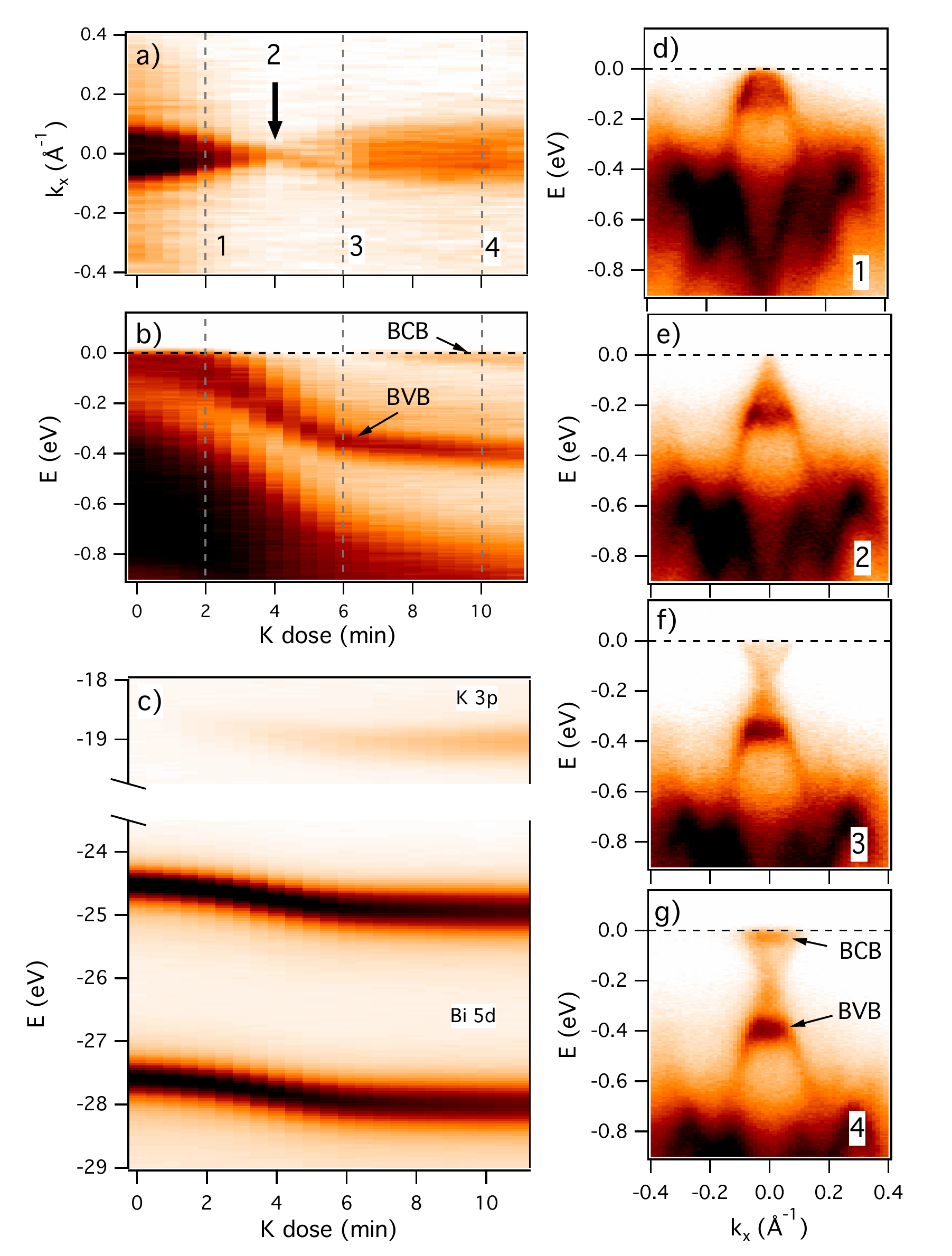}
\caption{Potassium doping of the Cr$_{0.15}$(Bi$_{0.1}$Sb$_{0.9}$)$_{1.85}$Te$_3$ surface. (a) Photoemission intensity at the Fermi level along the $\bar{M}-\bar{\Gamma}-\bar{M}$ momentum line as a function of K dose. The arrow (2) marks the neutrality point at which the originally $p$-type topological surface state turns into an $n$-type. (b) Photoemission intensity at $k_x=k_y=0$ ($\bar{\Gamma}$ point) as a function of K dose. (c) Intensity of K 3$p$ and Bi 5$d$ core levels vs K dose time. (d-g) Electronic dispersions along the $\bar{M}-\bar{\Gamma}-\bar{M}$ momentum line for the four specific K-doses (2, 4, 6 and 10 min), as indicated in (a). The positions of bulk valence band (BVB) and bulk conduction band (BCB) are marked in (b) and (g)
}
\label{Fig2}
\end{center}
\end{figure*}

\subsection{Pristine Cr$_{0.15}$(Bi$_{0.1}$Sb$_{0.9}$)$_{1.85}$Te$_3$}

Figure \ref{Fig1} shows the electronic structure of pristine, \textit{as-grown} Cr$_{0.15}$(Bi$_{0.1}$Sb$_{0.9}$)$_{1.85}$Te$_3$. Panel (a) represents the photoemission intensity from the narrow energy window around the Fermi level ($\pm$3 meV) as a function of in-plane momentum, representing the Fermi surface. Panels (b) and (c) show the electronic state dispersions along the $\bar{M}-\bar{\Gamma}-\bar{M}$ and $\bar{K}-\bar{\Gamma}-\bar{K}$ momentum lines, respectively, as indicated in (a). Panel (d) represents the photoemission intensity at the Fermi level on the larger momentum scale of several Brillouin zones to indicate that the states around the zone center, shown in panel (a), are the only states forming the Fermi surface. The linearly dispersing states (panels b and c) cross the Fermi level and form the hole-like, nearly circular Fermi surface. This is the lower portion of topological surface state (TSS). By extrapolating the occupied part of its dispersion, we estimate that the Dirac point is roughly $E_D\approx215$ meV above the Fermi level. This is significantly higher than what was found in STM study, where $E_D\approx165$ meV \cite{Lee2015c}. The TSS dispersion is also steeper than in STM studies, with the Fermi velocity varying slightly around the Fermi surface from $\sim3.5$ to $\sim4.4$ eV\AA$^{-1}$. The reason for these discrepancies could be in the variation of the Cr doping between the two samples used in this and STM study.
\begin{figure*}[htbp]
\begin{center}
\includegraphics[width=14cm]{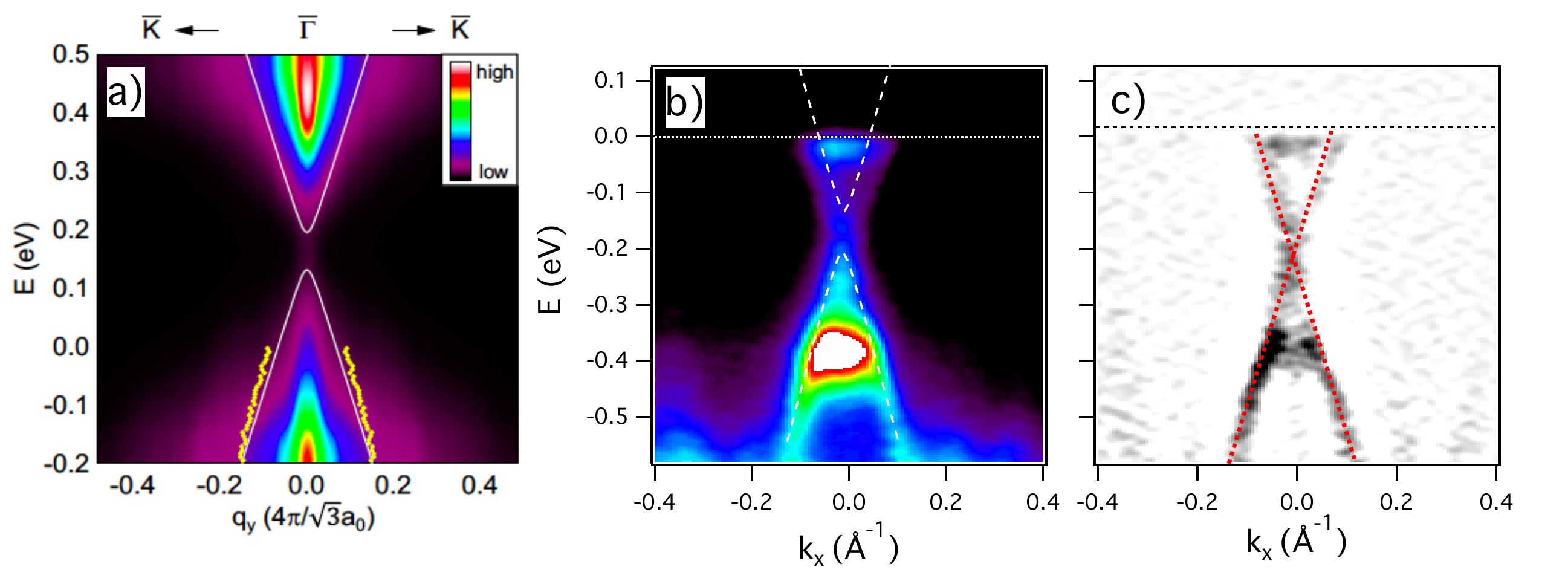}
\caption{Dirac cone in Cr$_{0.15}$(Bi$_{0.1}$Sb$_{0.9}$)$_{1.85}$Te$_3$. (a) STM quasiparticle interference spectra from the \textit{as grown} Cr$_{0.15}$(Bi$_{0.1}$Sb$_{0.9}$)$_{1.85}$Te$_3$ from ref. \cite{Lee2015c}. (b) ARPES spectrum along the $\bar{M}-\bar{\Gamma}-\bar{M}$ line in the SBZ from the K-doped sample after the 10 min dose where the  bulk conduction band is already partially occupied. (c) The second derivative in momentum direction, $d^2/(dk)^2$, of the ARPES spectrum from (b). The white dashed curves in (b) represent the same presumed dispersion of TSS with Dirac gap of 70 meV from (a). The red dotted lines in  (c) represent the gapless TSS Dirac cone. The STM data were taken at $\sim4.5$ K. The ARPES data were taken at 12 K.
}
\label{Fig3}
\end{center}
\end{figure*}

Aside form the surface Dirac cone, bulk valence states are also visible in Fig. \ref{Fig1}. The states forming the valence band maximum overlap with the TSS near the Fermi level, so the central nearly circular contour in Fig. \ref{Fig1}(a) is composed of both bulk and surface states. In addition to that, the top of the bulk valence band shows the three-fold symmetric protrusions along the three $\bar{\Gamma}-\bar{M}$ lines that just touch the Fermi level at the used photon energy. 
 
\subsection{Potassium Doped Cr$_{0.15}$(Bi$_{0.1}$Sb$_{0.9}$)$_{1.85}$Te$_3$}
Since the Dirac point is unoccupied in the pristine sample, the conventional ARPES is not able to probe it. Therefore, we have performed \textit{in-situ} surface electron doping by depositing potassium on the cleaved pristine sample. The results are shown in Fig. \ref{Fig2}. Panel (a) shows the intensity at the Fermi level along the $\bar{M}-\bar{\Gamma}-\bar{M}$ momentum line. It is apparent that upon electron doping, the initially hole like Fermi surface shrinks and at $\sim4$ min of K-dose reaches its minimum, after which it turns into an electron like one that grows in size with the additional K-dosing. The minimum corresponds to the neutrality point - the point in doping where the Fermi level reaches the Dirac point of TSS. With even higher dosing ($\sim7$ min), the bulk conduction band starts to get filled. This doping development is further illustrated in a sequence of spectra shown in panels (d-g). As more electrons are doped into the TSS, the Dirac point moves from being above the Fermi level, panel (d), to the Fermi level, panel (e), and finally to being occupied, panels (f-g). The energy shifts of valence/conduction bands as well as of the K 3$p$ and Bi 5$d$ core levels are illustrated in panels (b-c). It is also evident that the shifts are nearly linear in the range 2-6 min, when the TSS is the only state that is being filled, while the dependence is slower outside of that range, when in addition to TSS, the bulk valence and conduction bands are being filled. 

The important observation is that, aside from the shift in energy, the Dirac cone of the TSS remains otherwise unaffected by K deposition. The final result is the spectrum with both the valence and conduction bulk bands clearly visible and Dirac cone of the TSS spanning the bulk gap. Importantly, the Dirac cone is gapless, in stark contrast to the STM study \cite{Lee2015c}. Additionally, the absence of any discontinuity in the K-dosing in the 2-6 min. range in Fig. \ref{Fig2}(a-b) would argue against a significant Dirac gap. The ARPES spectra were recorded at 12 K, below the bulk Curie temperature, a regime where the Dirac cone is expected to be gapped by time-reversal symmetry breaking, but well above the sub-Kelvin temperature where QAHE was detected. The STM measurements were done at lower temperature ($\sim4.5$ K), but still far away from the QAHE regime. So, the question is - why there is such a big difference between the ARPES and STM results.

\subsection{Comparison of STM and ARPES Data}

In attempting to answer this question, we first show the side-by side comparison of the STM and ARPES data in Fig. \ref{Fig3}. Panel (a) shows the quasiparticle interference (QPI) data along the $\bar{K}-\bar{\Gamma}-\bar{K}$ line in the $q$ space from the pristine sample \cite{Lee2015c}. Panel (b) shows our ARPES data of the sample doped to the level where the bulk conduction band is partially occupied (10 min of K-dosing), while (c) represents the second derivative of the same spectrum in the $k$ direction. The white curves in (a) represent a presumed dispersion of the TSS with the Dirac gap of 70 meV. As evident from panels (b) and (c), this same dispersion does not describe the ARPES data well. The gapless Dirac cone, as indicated in panel (c), better describes the dispersion of TSS. 

We note that QPI is usually dominated by the elastic backscattering of electronic quasiparticles on impurities, where an electron from the state $E$, $\vec{k}$ is scattered to $E$, $\vec{-k}$. Therefore, in conventional materials with nearly circular constant electronic energy contour, the radius of a QPI contour should be double that of electron\rq{}s: $q=2k$ \cite{Crommie1993}. In TIs, however, the $\vec{k}$ and $\vec{-k}$ states have the opposite spins and the backscattering on non-magnetic impurities is forbidden. That is why the QPI is generally profoundly reduced for isotropic TSSs. For example, in Bi$_2$Se$_3$ the QPIs are essentially completely absent and determination of energy-momentum dispersions required Landau level spectroscopy in magnetic field \cite{Hanaguri2010}. In the case of Cr-doped samples, the back-scattering channel could partially re-open as dopands carry magnetic moments. However, the controlled ARPES studies found no difference in electronic scattering rates between the magnetic and non-magnetic impurities \cite{Valla2012a}. Therefore, one of the obvious reasons for dramatically reduced intensity of QPIs near the Dirac point in Cr$_{0.15}$(Bi$_{0.1}$Sb$_{0.9}$)$_{1.85}$Te$_3$ might be the suppression of backscattering. Another problem could be the contribution of bulk states - in ARPES, they can be easily distinguished, but in STM their contribution cannot be straightforwardly disentangled and could lead to erroneous assignment of different spectral features in TIs. For example, in topological crystalline insulator Pb$_{1-x}$Sn$_x$Se, there is no significant difference in conductance spectra of topologically trivial, fully gapped samples and those from the critical and even topological samples, illustrating that bulk states dominate those conductance spectra \cite{Gyenis2013,Zeljkovic2015}. The differences become visible in the external magnetic field when the Landau levels start to appear \cite{Zeljkovic2015}. The third reason could be that in both STM and ARPES, the high concentration of impurities leads to scattering rates being so high that they mask the real physics around the Dirac point that might be at a much smaller scale, comparable to the temperature range of QAHE. Finally, it might be that the TSS is not   sensitive to the bulk magnetism as it might not be significantly coupled to the bulk states responsible for magnetic ordering \cite{Swatek2020}. This would go contrary to suggestions that surface ordering temperature could be actually significantly higher than the bulk \cite{Lee2015c}.

\section{Summary}

In conclusion, we presented high-resolution ARPES study of the electronic properties of pristine and potassium doped surface of Cr$_{0.15}$(Bi$_{0.1}$Sb$_{0.9}$)$_{1.85}$Te$_3$. Initially a $p$-type, the TSS was turned into an $n$-type after potassium adsorption, allowing spectroscopic observation of Dirac point. In contrast to the gapped surface state observed in STM measurements, we found, within our detection limits, a gapless Dirac cone in this material. This suggests that the Dirac gap might be much smaller in magnitude and limited only to the regime where the QAHE was observed. In magnetically doped TIs, the scattering rates might be too high for the clear spectroscopic observation of a Dirac gap in these materials. The hope is now that in a much cleaner, intrinsic system, MnBi$_2$Te$_4$, in which the QAHE has been observed at significantly higher temperatures (5-10 K), the spectral features describing the Dirac cone would be clearer and reproducible. However, the ARPES studies give conflicting results, with strong indications that the Dirac gap is below the detection limits even in this material \cite{Otrokov2019,Zeugner2019,Rienks2019,Shikin2020,Hao2019,Chen2019b,Li2019,Swatek2020,Nevola2020}.

\section*{acknowledgments}
This work was supported by the US Department of Energy, Office of Basic Energy Sciences, contract no. DE-SC0012704. The Advanced Light Source was supported by the U.S. Department of Energy, Office of Sciences, under contract DE-AC02-05CH11231.

\section*{DATA AVAILABILITY}
The data that support the findings of this study are available from the corresponding author upon reasonable request.

\section*{references}
\bibliographystyle{ieeetr}

\end{document}